# Towards a new approach of extracting and querying fuzzy summaries

Ines Benali Sougui[1], Minyar Sassi Hidri[2], Amel Grissa Touzi[3]

[1,2,3]Université Tunis El Manar
[1,2]Ecole Nationale d'Ingénieurs de Tunis
[1,2]Laboratoire Signal, Images et Technologies de l'Information, Tunisia
[3]Faculté des Sciences de Tunis
[3]Laboratoire d'Informatique, Programmation, Algorithmique et Heuristiques, Tunisia

**ABSTRACT**

Diversification of DB applications highlighted the limitations of Relational Database Management System RDBMS particularly on the modeling plan. In fact, in the real world, we are increasingly faced with the situation where applications need to handle imprecise data and to offer a flexible querying to their users. Several theoretical solutions have been proposed. However, the impact of this work in practice, remained negligible with the exception of a few research prototypes based on the formal model GEFRED.

In this work, we proposed a new approach for exploitation of Fuzzy Relational Databases (FRDB) described by the model GEFRED. This approach consists of 1) A new technique for extracting summary fuzzy data, Fuzzy SAINTETIQ, based on the classification of fuzzy data and Formal Concepts analysis; 2) An approach of assessing flexible queries in the context of FDB based on the set of fuzzy summaries generated by our Fuzzy SAINTETIQ system; 3) An approach of repairing and substituting unanswered query.

Keywords: Fuzzy databases, Fuzzy summaries, GEFRED, Fuzzy-SaintEtiq, Fuzzy FCA, Flexible querying, Substituting query, Repairing query.

**INTRODUCTION**

Techniques based on data summaries are now considered as a good way to handle large amounts of data, especially when the precise values of these data are not needed. Thus, several databases (DB) summarization model have been proposed such as SaintEtiQ model (Raschia, 2002), which is close to our research tasks. This system makes it possible to generate a hierarchy of summaries making it possible to cover parts of the data base.

Unfortunately, all of these techniques are limited to the exact data and cannot be applied to fuzzy data.

We are confronted more and more with the situation where applications need to manage fuzzy data and to make profit their users from flexible querying. We speak then about flexible querying and Fuzzy Databases (FDB) (Galindo et al., 2006; Ben Hassine et al., 2008).

Among the results of the query, the user is more interested in the most important or the best, called the Top-K answers. The main reason for this interest is that they avoid overloading the user with a large number of uninteresting answers.

The querying mechanism also provides the means to detect the failed queries reasons. The purpose of the repairing approaches is to provide satisfactory answers even when the query does not accept a result in the strict sense. However, existing approaches only handle Boolean queries and cannot be applied to FRDB.

In this work, we present an extension of the SaintEtiQ summarization model for modeling fuzzy data and therefore to exploit to flexible FSQL query by turning the top *k* result. Our goal is also to found a result even in the case of the absence of summaries corresponding strictly to the query.

The rest of the paper is organized as follows: section 2 presents the basic concepts of Fuzzy Database (FDB), fuzzy attributes in GEFRED model and the basic concept of fuzzy FCA. Section 3 exposes the fuzzy queries' modeling and an overview of FCA-based summary model. Section 4 gives the limits of the existing summarization approach and of the exploitation of fuzzy data. Then it presents the architecture of our proposed approach. Section 5 describes the first step data organization of our proposed approach. Section 6 describes the second step the querying phase of our proposed approach. Section 7 gives a new kind of repaired query. Section 8 concludes the paper and gives some future work.

## BACKGROUND

In this section, we present the basic concepts of FDB, Fuzzy attributes in GEFRED model and the theoretical foundations of fuzzy FCA.

### Fuzzy Databases (FDB)

In this section, we present the basic concepts of FDB. A FDB is an extension of the relational database. This extension introduces fuzzy predicates under shapes of linguistic expressions that, at the time of a flexible querying, permits to have a range of answers (each one with a membership degree) in order to offer to the user all intermediate variations between the completely satisfactory answers and those completely dissatisfactory (Bosc et al., 1998). The FRDB models are considered in a very simple shape and consist in adding a degree, usually in the interval [0,1], to every tuple. It allows maintaining the homogeneity of the data in DB. The main models are those of PradeTestemale (Prade et al., 1987), Umano-Fukami (Umano et al., 1980), Buckles-Petry (Buckles et al., 1982), ZemankovaKaendel (Zemankova-Leech, 1985) and GEFRED of Medina et al. (Medina et al., 1994). This last model constitutes an eclectic synthesis of the various models published so far with the aim of dealing with the problem of representation and treatment of fuzzy information by using relational DB.

### Fuzzy attributes in GEFRED Model

The GEFRED model (GEneralised model Fuzzy heart Relational Database) has been proposed in 1994 by Medina et al. (Medina et al., 1994). One of the major advantages of this model is that it consists of a general abstraction that allows for the use of various approaches, regardless of how different they might look. In fact, it is based on the generalized fuzzy domain and the generalized fuzzy relation, which include respectively classic domains and classic relations. In order to model fuzzy attributes we distinguish between two classes of fuzzy attributes: fuzzy attributes whose fuzzy values are fuzzy sets and fuzzy attributes whose values are fuzzy degrees (Galindo, 2005; Galindo et al., 2006).

*Fuzzy Sets as Fuzzy Values.* These fuzzy attributes may be classified in four data types. This classification is performed taking into account the type of referential or underlying domain. In all of them the values Unknown, Undefined, and Null are included:

- Fuzzy Attributes Type 1 (FTYPE1): These are attributes with "precise data", classic or crisp (traditional, with no imprecision). However, they can have linguistic labels defined over them, which allow us to make the query conditions for these attributes more flexible.

- Fuzzy Attributes Type 2 (FTYPE2): These attributes admit both crisp and fuzzy data, in the form of possibility distributions over an underlying ordered domain (fuzzy sets). It is an extension of the FTYPE1 that does, now, allow the storage of imprecise information.
- Fuzzy Attributes Type 3 (FTYPE3): They are attributes over "data of discreet non-ordered dominion with analogy".
  In these attributes some labels are defined ("blond", "red", "brown", etc.) that are scalars with a similarity (or proximity) relationship defined over them, so that this relationship indicates to what extent each pair of labels be similar to each other.
- Fuzzy Attributes Type 4 (FTYPE4): These attributes are defined in the same way as Type 3 attributes, without it being necessary for a similarity relationship to exist between the labels.

*Fuzzy Degrees as Fuzzy Values.* The domain of these degrees can be found in the interval [0,1], although other values are also permitted, such as a possibility distribution (usually over this unit interval) (Medina et al., 1994), (Galindo, 2005). The meaning of these degrees is varied and depends on their use. The most important possible meanings of the degrees used by some authors are: Fulfillment degree, Uncertainty degree, Possibility degree and Importance degree.

## Fuzzy FCA

In this section, we discuss the Fuzzy FCA proposed by (Quan et al., 2004), which incorporates fuzzy logic into Formal Concept Analysis, to represent vague information.

**Definition 1.** A fuzzy formal context is a triple $K_f = (G, M, I = \varphi(G \times M))$ where $G$ is a set of objects, $M$ is a set of attributes, and $I$ is a fuzzy set on domain $G \times M$. Each relation $(g,m) \in I$ has a membership value $\mu(g,m)$ in $[0, 1]$.

A fuzzy formal context can also be represented as a cross-table as shown in table 1. The context has three objects representing three documents, namely *D1, D2* and *D3*. In addition, it also has three attributes, *Data Mining* (*D*), *Clustering* (*C*) and *Fuzzy Logic* (*F*) representing three research topics. The relationship between an object and an attribute is represented by a membership value between 0 and 1.

A confidence threshold *T* can be set to eliminate relations that have low membership values (Quan et al., 2004). Table II shows the cross-table of the fuzzy formal context given in table 1 with $T = 0.5$.

*Table 1. Fuzzy Formal Context with T = 0.5*

|    | D   | C    | F    |
|----|-----|------|------|
| D1 | 0.8 | 0.12 | 0.61 |
| D2 | 0.9 | 0.85 | 0.13 |
| D3 | 0.1 | 0.14 | 0.87 |

Each relationship between the object and as a membership value in fuzzy formal context, then the intersection of these membership values should be the minimum of these membership values, according to fuzzy theory (Zadeh, 1965).

**Definition 2.** Fuzzy formal concept : Given a fuzzy formal context $K_f = (G, M, I = \varphi(G \times M))$ and a confidence threshold *T*, we define $A* = \{m \in M | \forall g \in A: \mu(g, m) \geq T\}$ for $A \subseteq G$ and $B* = \{g \in G | \forall m \in B: \mu(g,m) \geq T\}$ for $B \subseteq M$.

A fuzzy formal concept (or fuzzy concept) of a fuzzy formal context $K_f = (G, M, I = \varphi(G \times M))$ with a confidence threshold $T$ is a pair $(A_f = \varphi(A), B)$ where $A \subseteq G$, $B \subseteq M$, $A* = B$ and $B* = A$. Each object $g \in \varphi(A)$ has a membership $\mu_g$ defined as $\mu_g = min(\mu(g,m))$ and $m \in B$ where $\mu(g,m)$ is the membership value between object $g$ and attribute $m$, which is defined in $I$. Note that if $B = \{\}$ then $\mu_g = 1$ for every $g$.

**Definition 3.** Let $(A_1, B_1)$ and $(A_2, B_2)$ be two fuzzy concepts of a fuzzy formal context $(G, M, I)$. $(\varphi(A_1), B_1)$ is the sub-concept of $(\varphi(A_2), B_2)$, denoted as $(\varphi(A_1), B_1) \leq (\varphi(A_2), B_2)$, if and only if $\varphi(A_1) \subseteq \varphi(A_2)(\Leftrightarrow B_2 \subseteq B_1)$.
Equivalently, $(A_2, B_2)$ is the super-concept of $(A_1, B_1)$.

**Definition 4.** A fuzzy concept lattice of a fuzzy formal context $K_f$ with a confidence threshold $T$ is a set $F(K_f)$ of all fuzzy concepts of $K_f$ with the partial order $\leq$ with the confidence threshold $T$.

**Definition 5.** The similarity of a fuzzy formal concept $K_1 = (\varphi(A_1), B_1)$ and its subconcept $K_2 = (\varphi(A_2), B_2)$ is defined as

$$E(K_1, K_2) = \frac{|\varphi(A_1) \cap \varphi(A_2)|}{|\varphi(A_1) \cup \varphi(A_2)|} \quad (1)$$

Figure 1 (a) gives the traditional concept lattice generated from table 1, in which crisp values *Yes* and *No* are used instead of membership values. Figure 1 (b) gives the fuzzy concept lattice generated from the fuzzy formal context given in table 1.

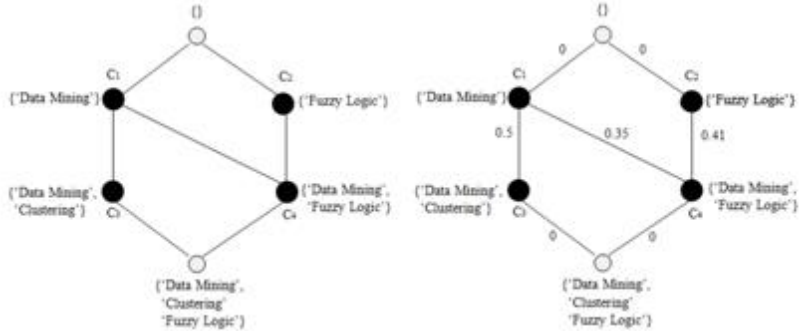

*Figure 1. (a) A concept lattice generated from traditional FCA. (b) A fuzzy concept lattice generated from Fuzzy FCA*

## RELATED WORK

In this section, we present the fuzzy queries modeling and an overview of FCA-based summary model (Sassi et al., 2010).

### Fuzzy queries' modeling

To model flexible queries, Medina et al. (Medina et al., 1998) introduces an extension of the SQL language called FSQL (Fuzzy SQL). So all valid queries in SQL are also valid in FSQL. This Flexibility comes from the fact that we can introduce fuzzy attributes, Fuzzy constants, fuzzy comparators, fuzzy qualifiers and quantifiers, etc. while returning a range of responses each with a satisfaction level. Each atomic condition combines a satisfaction level µϵ [0, 1] with an attribute value. For all the attributes of an n-uplet, the semantics of degrees are the same which involve that all criteria are commensurable. FSQL query has the following syntax:

> *Select* < attribute >
> *From* < strict relation >
> *Where* < fuzzy condition > THOLD α

where < fuzzy condition > can incorporate blocks of queries nested or partitioned. Parameters α from *Select* clause limit the number of answers by using a qualitative calibration (the data that satisfy the query with a threshold greater than α).

The FSQL language enriches the SQL language with:
- Linguistic labels: If an attribute is accessible to fuzzy processing, then linguistic labels can be defined on it, preceded by the $ symbol to distinguish them easily. These labels are of two types, depending on the attribute domains, ordered (used for Ftype 1 and 2) or unordered (used for Ftype 1, 3, and 5).
- Fuzzy comparators: In addition to the usual comparators (=,>, etc.), FSQL includes fuzzy comparators that can compare a column (or attribute) to a constant or two columns of the same type. These comparators are presented in table 2.

*Table 2. Fuzzy comparators*

| Id | Name | Age |
|---|---|---|
| **FEQ** | NFEQ | Fuzzy Equal |
| **FGT** | NFGT | Fuzzy Greater Than |
| **FGEQ** | NFGEQ | Fuzzy Greater or Equal |
| **FLT** | NFLT | Fuzzy Less Than |
| **FLEQ** | NFLEQ | Fuzzy Less or Equal |
| **MGT** | NMGT | Much Greater Than |
| **MLT** | NMLT | Much Less Than |

- The fuzzy qualifiers: they are of two natures, absolute and relative:
  1. Absolute qualifiers can answer such questions *the total number of tuples is large, small, much, a little, etc.* In this case, the qualifier depends on a single quantity.
  2. Relative qualifiers can answer questions in which the truth of the qualifiers depends on two quantities.
- Fuzzy constants: In addition to the usual constants, FSQL includes fuzzy constants (UNKNOWN, UNDEFINED, NULL, …).
- Fuzzy attributes: The classification adopted in the GEFRED model for attribute types is based on "imprecise" representation and data processing approaches.

With FSQL language, preferences are considered only as constraints and are taken into account through the expression of fuzzy predicates commensurable, modeled using fuzzy sets of values more or less satisfactory.

Besides, Top-k queries have attracted much interest in many different areas such as network and system monitoring (Babcock et al., 2003), information retrieval (Kimelfeld et al., 2006), sensor networks (Silberstein et al., 2006), (Wu et al., 2006), large databases (Grissa-Touzi et al., 2012), multimedia databases (Chaudhuri et al., 2004), spatial data analysis (Ciaccia et al., 2002), (Hjaltason et al., 2003), P2P systems (Akbarinia et al., 2007), data stream management systems (Mouratidis et al., 2006), etc.

The main reason for such interest is that they avoid overwhelming the user with large numbers of uninteresting answers which are resource-consuming.

The problem of answering top $k$ queries can be modeled as follows (Fagin et al., 2003). Suppose we have m lists of n data items such that each data item has a local score in each list and the lists are sorted according to the local scores of their data items. And each data item has an overall score which is computed based on its local scores in all lists using a given scoring function. Then the problem is to find the $k$ data items whose overall scores are the highest. This problem model is simple and general. Let us illustrate with the following examples. Suppose we want to find the top $k$ tuples in a relational table according to some scoring function over its attributes. To answer this query, it is sufficient to have a sorted (indexed) list of the values of each attribute involved in the scoring function, and return the $k$ tuples whose overall scores in the lists are the highest.

We plan to extend the FSQL language to support the top-k best answers. This new language is called FSQL-extended.

**The FCA-based Summary**

In (Sassi et al., 2010), they proposed to extend the SaintEtiQ summarization model (Raschia et al., 2002) by introducing some optimization processes including minimization of the expert risks domain, building of the summary hierarchy from DB records, and cooperation with the user by giving him summaries in different hierarchy levels. In this case, the summarization process can be divided into two major steps. For the first step, the pre-processing, we have considered a fuzzy clustering that permits the generating of a membership matrix associating the DB records to generated clusters through the membership degrees. The second step, the post processing, generates the summary hierarchy. Thus, we have proposed to use fuzzy FCA in order to generate fuzzy hierarchy. In this step, we cooperate with the user by giving him summaries in different hierarchy levels.

**SOLUTIONS AND RECOMMENDATIONS**

In recent years, several methods of DB summarization have been proposed. Unfortunately, all these techniques cannot be applied to the large Fuzzy DB, they can only applied to simple data sets.

In this paper, we propose a new approach of linguistic summarization for fuzzy databases, called Fuzzy-SaintEtiq (Benali-Sougui et al., 2013; Benali-Sougui, 2016). This approach is an extension of the SaintEtiQ summarization model (Raschia et al., 2002) and the FCA-based Summary (Sassi et al., 2010) to support the fuzzy data represented by GEFRED model.
Although our solution is based on the fuzzy model GEFRED and it can be applied to other fuzzy models.

In fact, we are confronted with the situation where applications need to manage fuzzy summaries and to profit their users from flexible querying. We have proposed a flexible FSQL query based on fuzzy linguistic summaries.

In the following we present the architecture of our proposed approach, which is divided in two major phases: fuzzy data organization phase and querying fuzzy summaries phase.

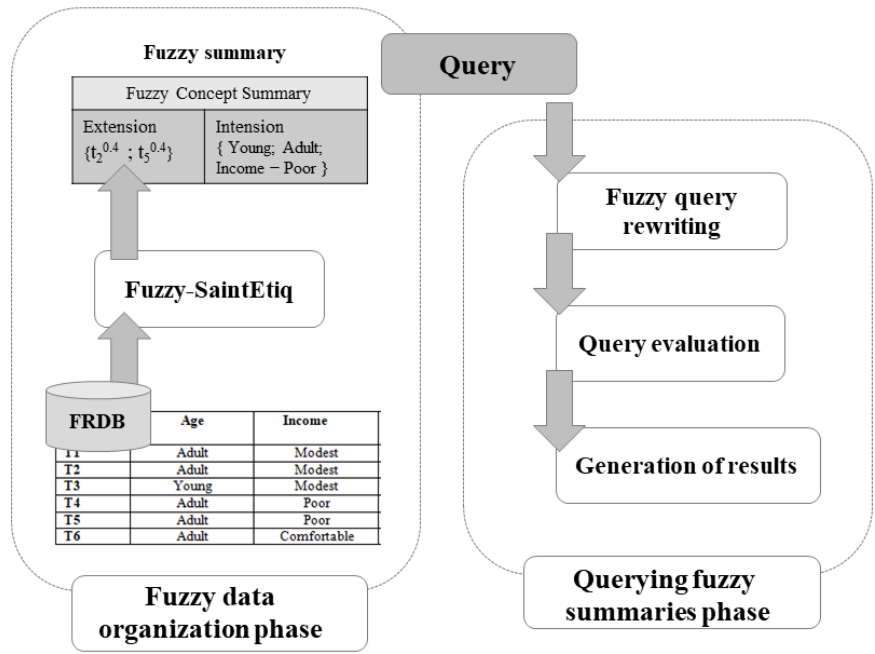

*Figure 2. The process of extracting and querying fuzzy summaries*

However, querying the summaries as explained above is interesting in order to rapidly get a rough idea of the properties of tuples in a relation but sometimes queries may have a null answer.

The user has to think about another query, which might fail. Thus, the idea is to repair and substitute query to provide an approximate answer.

Our goal in this paper is also to found a result even in the case of the absence of summaries corresponding strictly to the query. So, we speak about repaired query.

## FUZZY DATA ORGANIZATION PHASE

Fuzzy-SaintEtiq takes the FDB records and provides knowledge. The summarization act considered like a process of knowledge discovery from database, in the sense that it is organized according to two following principal steps.

- The preprocessing step: This step organizes the database records in homogeneous clusters having common properties. This step gives a certain number of clusters for each attribute. Each tuple has values in the interval [0..1] representing these membership degrees according the formed clusters. Linguistic labels, which are fuzzy partitions, will be assigned on attributes domain.
- The post treatment step: This step takes into account the result of the fuzzy clustering on each attribute, visualizes by using the fuzzy concepts lattices. Then, it imbricates them in a fuzzy nested lattice. Finally, it generalizes them in a fuzzy lattice associating all records in a simple and hierarchical structure. Each lattice node is a fuzzy concept which represents a concept summary.

  This structure defines summaries at various hierarchical levels. This step consists in organizing the summaries within a hierarchy such that the most general concept summary is placed at the root of the fuzzy lattice, and the most specific concepts summaries are the leaves. This summary model corresponds to prototypical approaches since the intention of a concept summary present for each attribute the various possible values in the form of a fuzzy descriptors and the representativeness of these descriptors within the specified concept summary.

**Fuzzy clustering**

For clustering fuzzy data, we have used an extension of Fuzzy C-Means clustering algorithm to deals with fuzzy data called Fuzzy-FCM (Grissa-Touzi, 2010; Benali-Sougui et al., 2013) in order to support different types of data represented by GEFRED model (Medina et al., 1994).

The Fuzzy-FCM algorithm allows the user to select attributes according to which he wants to carry out classification. This treatment gives a refined intermediate matrix only formed of the codes of the selected attributes. Once the selection achieved, the FCM algorithm is applied on the refined table to get a matrix of adherence and a cut is exercised on this matrix of adherence to purify it by eliminating all values lower to the cut. The main idea of the algorithm is to define an intermediate matrix to model fuzzy data.

Let consider a relation *R* = (*Id, Age, Income, Experience*) from an FDB Employee table. Table 3 gives a sample of FDB Employee table. In the following we will use this example.

*Table 3. Data sample*

| Id | Age | Income | ProfessionalBackground |
|---|---|---|---|
| **T1** | Adult | Modest | 10 |
| **T2** | Adult | Modest | 5 |
| **T3** | Young | Modest | 3 |
| **T4** | Adult | Poor | 20 |
| **T5** | Adult | Poor | 7 |
| **T6** | Adult | Comfortable | 12 |

Each cluster of a partition is labeled by linguistic descriptor provided by a domain expert. For example, the fuzzy label Young belongs to a partition built on the domain of age attribute. Linguistic variables associated with the attributes of *R*. These linguistic variables constitute the new attribute domains used for the rewriting of tuples in the summarization process. For clusters generation, we carry out a fuzzy clustering (Vanisri, 2010) while benefiting from fuzzy logic.
This operation makes it possible to generate, for each attribute, a set of membership degrees.

Figure 3 presents the results of fuzzy clustering applied to *Age and Income*.

For Age, Income and Experience attributes, fuzzy clustering generates three clusters. The minimal value (resp. maximal) of each cluster corresponds to the lower (resp.higher) interval terminal of the values of this last. Each cluster of a partition is labeled with a linguistic descriptor provided by the user or a domain expert. For this, the following abbreviations are used:

- For *Age* attribute: Young Age, Adult Age and Old Age.
- For *Income* attribute: Low Income, Modest Income and Comfortable Income.

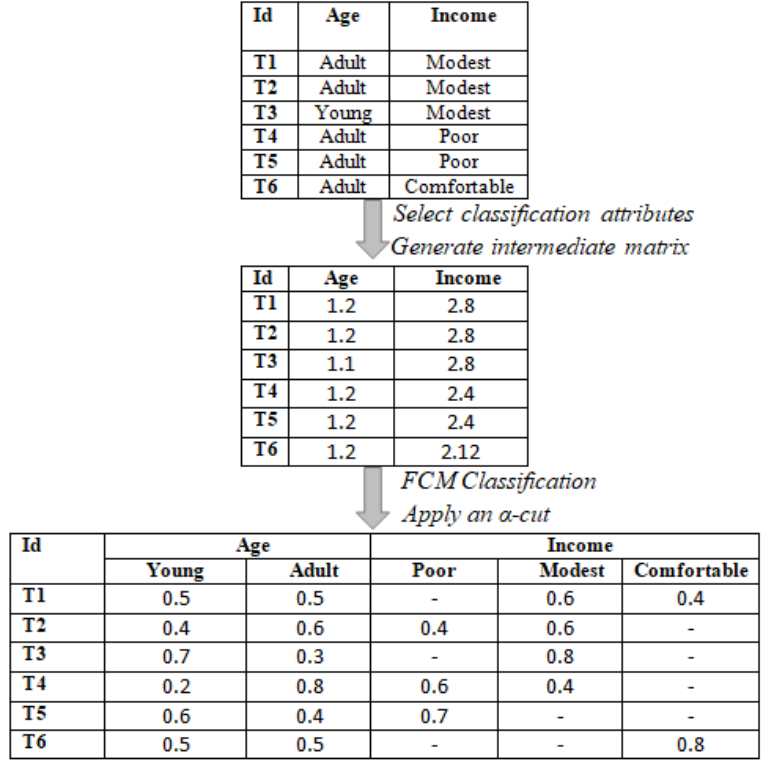

*Figure 3. Fuzzy-FCM process applied on the Employee table*

## Fuzzy summarization

After clustering, we generate the corresponding fuzzy concept lattice with similarity distance (Sassi et al., 2010).

For a more formal expression of a query, let be consider:
- $A = \{A_1,..., A_k\}$ is a set of attributes.
- $A_k$ is the $k^{th}$ attribute which appears is query $Q$,
- $R(A)$ is the relation whose tuples are summarized,
- $t_i$ is the $i^{th}$ tuple, $i = 1..N$,
- $L_k = \{l_{1k},..., l_{jk}\}$ is a set of linguistic terms of attribute $A_k$,
- $\mu_{ijk}$ is a membership degree of tuple $t_i$ (record) to the linguistic term $l_{kj}$ of attribute $A_k$.
- $R_{zf}$ is the sub set of involved tuples in summary $z_f$,
- $I_{zf}$ is the sub set of linguistic terms which appears in summary $z_f$,
- $z_f = (R_{zf}, I_{zf})$ is the concept summary, in which $R_{zf}$ and $I_{zf}$ are respectively the extension and the intension of the concept,
- $level$ is the level of the concept summary in the concept lattice,
- $R_z = \{t_1, t_2,..., t_N\}$. Each $t_i$ is associated to one primitive DB records, i.e. an element of $R$.

- $\left|R_{zf}\right|$ is the number of candidates tuples in $R_{zf}$.

Table 4 shows the fuzzy formal concept lattice generated according the fuzzy formal context given in figure 2.

Each concept summary $z_f$ of the set of concept's summary $Z_f$ is thus a description of a set of n-uplets of $R$ which jointly form his extension and which is noted by $R_{zf}$. Table 4 shows concept summaries for each level extracted from fuzzy concept lattice. Thus, the root contains the summary of all the candidate records, whereas leaves represent only one combination of fuzzy linguistic labels over all the attributes. Levels 0 and 5 are both the root and leaves concept summary.

*Table 4. Fuzzy Concept summaries' generation*

| Level | Fuzzy summaries |
|---|---|
| 0 | $z_0 = (\{t_1; t_2; t_3; t_4; t_5; t_6\}; \emptyset)$ |
| 1 | $z_{11} = (\{t_1^{0.5}; t_2^{0.6}; t_4^{0.8}; t_5^{0.4}; t_6^{0.5}\}; \{Adult\})$<br>$z_{12} = (\{t_1^{0.5}; t_2^{0.4}; t_3^{0.7}; t_5^{0.6}; t_6^{0.5}\}; \{Income - Modest\})$<br>$z_{13} = (\{t_1^{0.5}; t_2^{0.4}; t_3^{0.7}; t_5^{0.6}; t_6^{0.5}\}; \{Young\})$ |
| 2 | $z_{21} = (\{t_1^{0.5}; t_2^{0.6}; t_4^{0.4}\}; \{Adult; Income - Modest\})$<br>$z_{22} = (\{t_1^{0.5}; t_2^{0.4}; t_3^{0.7}\}; \{Young; Income - Modest\})$<br>$z_{23} = (\{t_2^{0.4}; t_4^{0.6}; t_5^{0.7}\}; \{Adult; Income - Poor\})$<br>$z_{24} = (\{t_1^{0.5}; t_{0.6}^2; t_5^{0.4}; t_6^{0.5}\}; \{Young; Adult\})$ |
| 3 | $z_{31} = (\{t_2^{0.4}; t_4^{0.6}\}; \{Adult; Income - Poor; Income - Modest\})$<br>$z_{32} = (\{t_1^{0.5}; t_2^{0.4}\}; \{Young; Adult; Income - Modest\})$<br>$z_{33} = (\{t_2^{0.4}; t_5^{0.4}\}; \{Young; Adult; Income - Poor\})$<br>$z_{34} = (\{t_1^{0.4}; t_6^{0.8}\}; \{Young; Adult; Income - Comfortable\})$ |
| 4 | $z_{41} = (\{t_2^{0.4}\}; \{Young; Adult; Income - Poor; Income - Modest\})$<br>$z_{42} = (\{t_1^{0.4}\}; \{Young; Adult; Income - Modest; Income - Comfortable\})$ |
| 5 | $z_5 = (\{\emptyset\}; \{Young; Adult; Income - Poor; Income - Modest; Income - Comfortable\})$ |

## QUERYING FUZZY SUMMARIES PHASE

This phase takes place in 3 steps which will be developed later:
- Fuzzy query rewriting
- Query evaluation
- Generation of results.

### Fuzzy query rewriting

In the FSQL-extended query we have two parameter *α* and *k*. As previously said parameters *α* and *k* from Select clause limits the number of answers by using a quantitative calibration (*k* best responses) or qualitative calibration (the data that satisfy the query with a threshold greater than *α*).

The parameter *k* is given by the user and the parameter is calculated depending on the number of clusters which *α* involved in condition of Select query.

$$\alpha = \frac{1}{\max(\text{NClus})} \quad (2)$$

where NClus is the number of clusters involved in the condition of select clause.

Let us consider the example in the table 3. The queries are as follows:

> **$Q_1$: What is the experience and income of each young employee with a minimum degree of 0.5.**

*Select* Income, ProfessionalBackground
*From* Employee
*Where* Age FEQ $Jeune THOLD 0.5;

And

> **$Q_2$: What is the experience of "young" employees with a "comfortable" income with a minimum degree of 0.3.**

*Select* ProfessionalBackground
*From* Employee
*Where* Age FEQ $Young THOLD 0.3 AND Income FEQ $Comfortable THOLD 0.3;

In a query, descriptors like Young, Comfortable in $Q_1$ and $Q_2$ are called required characteristics and embody the properties that a record must consider them as an element of the answers. A query also defines the attributes for which required characteristics exist. The set of these input attributes is denoted by Inputs($A_Q$). The expected answer is a description over a set of other attributes, denoted by Outputs($A_Q$). It is the complement of Inputs($A_Q$) relatively to $A_Q$ (the set of attributes appears in the query Q):

$$Inputs(A_Q) \cup Outputs(A_Q) = A \qquad (3)$$

and

$$Inputs(A_Q) \cap Outputs(A_Q) = \emptyset \qquad (4)$$

Hence a query $Q$ defines not only a set *Inputs($A_Q$)* of input attributes but also for each attribute $A_k$, the set $L_{Ak}(Q)$ of its required characteristics which define the set of linguistic terms of attribute $A_k$ appears query $Q$. The set of sets $L_{Ak}(Q)$ is denoted by $L(Q)$.

For example, for $Q_2$, this set is determined as follows:
- *Inputs($A_{Q2}$)* = {Income, Age};
- *Outputs($A_{Q2}$)* = { ProfessionalBackground };
- $L_{Age}(Q_2) = \{Young\}$, $L_{Income}(Q_2) = \{Comfortable\}$;
- $L(Q_2) = \{L_{Age}(Q_2), L_{Income}(Q_2)\}$.
- The degree of membership to this query is 0.3.

The query rewriting in a logical proposition $P_f(Z_f, Q)$ used to qualify the link between the fuzzy summary $Z_f$ and the query $Q$. $P_f(Z_f, Q)$ is in a conjunctive form in which all descriptors are literals. Then, each set of descriptors yields one corresponding clause. Thereafter we will apply an $\alpha$-cut on this new form with the parameters $\alpha$ is calculated previously.

Let be consider the query Q3:

> **Q3: Which are the "young" and "adult" employees who have a "poor" or "modest" income with a minimum degree of 0.3.**

*Select* *
*From* Employee
*Where* Age FEQ ($Young, $Adult) THOLD 0.3;
*And* Income FEQ ($Poor, $Modest) THOLD 0.3;

We have then:

- *Inputs(A_Q3)* = {*Age, Income*};
- $L_{Age}$ = {*Young, Adult*};
- $L_{Income}$ = {*Poor, Modest*};
- The degree of membership to this query is 0.3;
- P(Q3) = (0,3-*cut*(*Young*) ∨ 0,3-*cut* (*Adult*)) ∧ (0,3-*cut* (*Poor*) ∨ 0,3-*cut*(*Modest*)).

## Query evaluation

In this section, we try to evaluate the proposed approach. For this, the searching procedure should take into account all fuzzy summaries in the concept lattice that correspond to the query *Q*.

Our approach is based on our definition of fuzzy summary α-summary and the valuation function, which are defined as follows:

**Definition 6.** (BenAli-Sougui et al. 2013) An *α-summary*, denoted as $Z_f$, is a fuzzy summary $Z_f$ = ($R_{Zf}$, $I_{Zf}$) in which $Z_f$ is a collection of candidate records $R_{Zf}$ = {$t_1, t_2, ... ,t_N$} which represents the extent and $I_{Zf}$ is the intent. Each tuple $t_i$ of $R_{Zf}$ existing in the summary has a membership value. It can be formulated as follows:

$$Z_f = \{ \forall t_i \in Z_f | \mu(t_i) \geq \alpha \} \text{ with } \mu(t_i) \in [0, 1] \quad (5)$$

**Definition 7.** (BenAli-Sougui et al. 2014) Let be consider $v_f$ the valuation function. It is obvious that the valuation of $P_f(Q)$ depends on the summary $Z_f$. Thus $v_f(P_f(Q)_{Zf})$ denotes the valuation of $P_f(Q)$ in the context of $Z_f$. $L_{Ai}(Z_f)$ the set of linguistic terms that appear in $Z_f$ and $L_{Ai}(Q)$ the set of linguistic terms that appear in query *Q*. We can distinguish between three assumptions:

- *Coresp*($Z_f$, *Q*) = *Exact* : $v_f(P_f(Q)_{Zf})$ = *true* and $L_{Ai}(Z_f) \subseteq L_{Ai}(Q)$ : All tuples including in $Z_f$ verify the query *Q*;
- *Coresp*($Z_f$, *Q*) = *False* : $v_f(P_f(Q)_{Zf})$ = *false* and $L_{Ai}(Z_f) \neq L_{Ai}(Q)$ Linguistic terms appear in $Z_f$ do not correspond to terms in query Q;
- *Coresp*($Z_f$, *Q*) = *Indecision* : ∃i, $L_{Ai}(Z_f) - L_{Ai}(Q) \neq \emptyset$: There are some tuples in $Z_f$ satisfying *Q*.

These situations reflect a global view of the matching of a fuzzy summary $Z_f$ with a query *Q*.

Figure 4 presents the steps of our flexible query approach based on fuzzy data summaries.

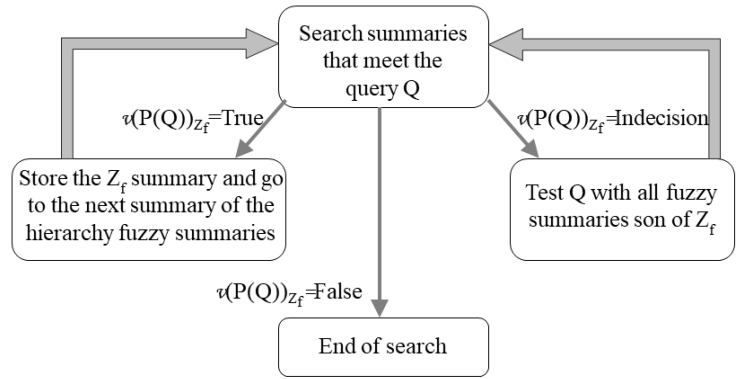

*Figure 4. The flexible querying approach of fuzzy summaries*

Algorithm 1 describes the different steps of our approach.

**Algorithm 1** SearchResult ($Z_f$, level, Q)

**Require:** $Z_f$ the fuzzy concept summary, *level* the current level of summary, Q the query.
**Ensure:** $L_{result}$ list of all fuzzy summary responding to the query Q.
1: $L_{result} \leftarrow \phi$
2: **If** ($Coresp(Z_f, Q) = Exact$) **then**
3:    $Insert(Z_f, L_{result})$
4: **Else**
5:    **If** ($Coresp(Z_f, Q) = indecision$) **then**
6:       **For all** Fuzzy summary $z_f$ son of $Z_f$ **do**
7:          level $\leftarrow$ level+1
8:          $L_{result} \leftarrow L_{result}$+ SearchResult ($Z_f$, level, Q)
9:       **End for**
10:    **End if**
11: **End if**
12: **Return** $L_{result}$

The procedure $Insert(Z_f, L_{result})$ lets us to insert $Z_f$ in the list $L_{result}$ of all fuzzy summary responding to the query Q.
$Coresp(Z_f, Q)$ is the function that lets to test the correspondence between the summary $Z_f$ and the query Q that we have seen previously.

The idea of our approach is to search using the summary concept; which summary responds to the query and calculate their satisfaction degree. Then we will insert them in a list order by satisfaction degree. We will repeat these steps until ensure that there is not a branch in the summary concept that satisfies the query. Finally we can display the top *k α-summary* from the list of results. Figure 5 shows the principal of our approach.

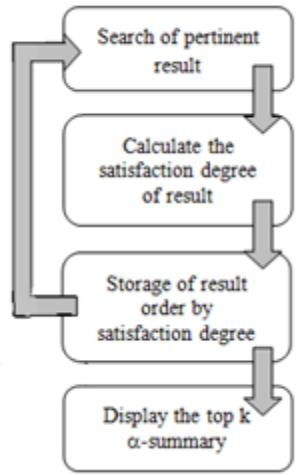

*Figure 5. The steps of the flexible querying approach of the top-k fuzzy data summaries*

The satisfaction degree SD of summary $Z_f$ is calculated by a function that returns the best route that lets us to find $Z_f$. It is determined as follows:

$$SD = \max(\sum_{j=1}^{p} Fuzzy\_score(Zj, Zj-1)) \qquad (6)$$

with p the current level of $Z_f$

$$\text{Fuzzy\_score}(Zp, Zp-1) = \frac{|\varphi(Zp) \cap \varphi(Zp-1)|}{|\varphi(Zp) \cup \varphi(Zp-1)|} \quad (7)$$

The equation of Fuzzy-score is proposed by (T.T. Quan, 2004).

## Generation of results

Recall that the hierarchy of summary concepts in our Employee test table is presented in table 3.

The result of applying the proposed approach on the concept lattice for queries $Q_1$, $Q_2$ and $Q_3$ for k = 3 is given in table 6.

*Table 6. Top k α-summary result*

| Query | α-Summary |
|---|---|
| Q1 | $\alpha - z_{13} = \{t_1^{0.5}, t_3^{0.7}, t_5^{0.6}, t_6^{0.5}\}$ |
| Q2 | $\alpha - z_{42} = \{t0.41\}, \alpha - z_{34} = \{t0.41, t0.86\}$ |
| Q3 | $\alpha - z_{21} = \{t_1^{0.5}, t_2^{0.6}, t_4^{0.4}\}, \alpha - z_{22} = \{t_1^{0.5}, t_2^{0.4}, t_3^{0.7}\}, \alpha - z_{23} = \{t_2^{0.4}, t_4^{0.6}, t_5^{0.7}\}$ |

## REPAIRING QUERY

Our objective is to found a result even in the case of the absence of summaries corresponding strictly to the query.

It is thus possible, in an interactive mode, to present to the user the reasons for the failure of his query and to submit alternative queries. To meet this goal, we will present our approach to extracting and querying fuzzy summaries with repairing unanswered query (BenAli-Sougui et al. 2014) described in figure 6.

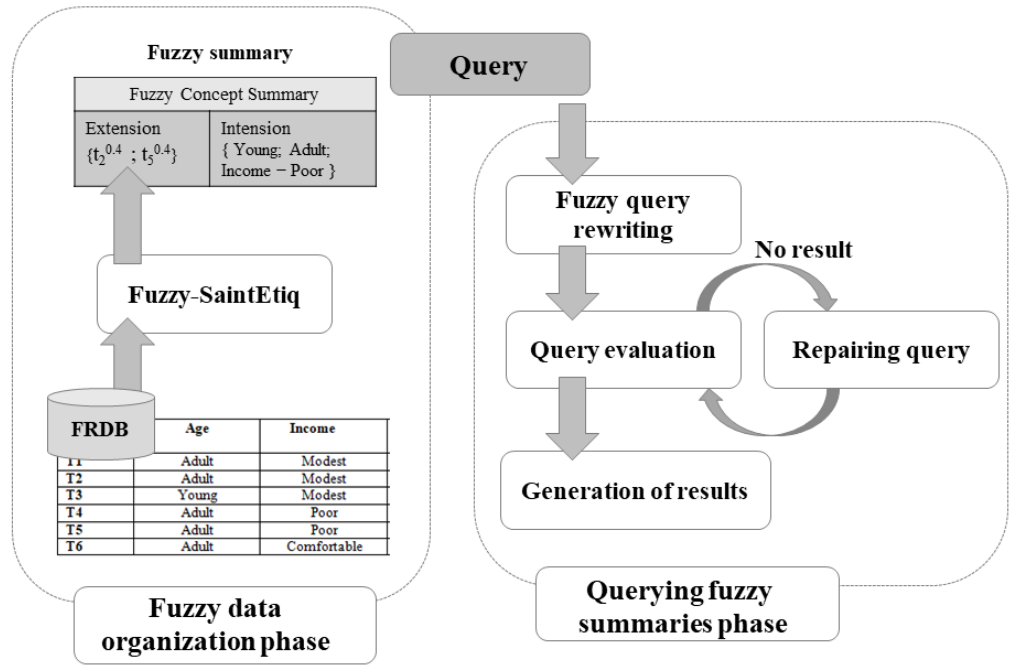

*Figure 6. The process of extracting and querying fuzzy summaries with unanswered query repair*

To accommodate this target we will present a new repaired query. Reparation is based on the idea that there could be a result semantically close to the query. To find these results, the query is modified using the best fuzzy summaries which is the near value answering the query.

Repairing queries has already been implemented in the context of a mediator by Bidault et al. (Bidault et al., 2000), as one of the cooperative aspects reviewed by Gaasterland et al. in (Gaasterland et al., 1992).

**Basic principal**

This process happens each time a tree exploration that is conducted to find answers for a query $Q$ fails at a specific summary $Z_f$ (i.e., when some required characters are absent from $Z_f$). $Z_f$ is then called a failure node. As the exploration may fail for more than one summary, several modifications of the same query can be proposed to the user.

The strategy is guided by the fuzzy summary hierarchy and it guarantees that results will be found for the new query.

For short, repairing query principe follows these steps:

**Step 1:** Evaluate the initial query $Q$;

**Step 2:** If no result, execute the substitution of query;

**Step 3:** Determine the best substitution query $Q_{sub}$, if the son of $Z_f$ have other values for the attributes who has failed then calculate the distance measure between $Q$ and sons of $Z_f$ and use the attributes of the son of $Z_f$ which has the greatest distance value, else use the attributes values of $Z_f$;

**Step 4:** Evaluate $Q_{sub}$;

**Step 5:** Express the results of $Q_{sub}$.

More distance value is greater, more the summary and query have common attributes.

The distance measure is calculated as follows:

**Definition 7.** The distance between a query $Q$ and summary $Z_f$, denoted as $D_Q(Z_f)$, is the sum of the number of common attributes appears in the query $Q$ and summary $Z_f$.

$L_{A_k}$ is the set of linguistic terms of attribute $A_k$ appears in the query $Q$ and $I_{Z_f}$ is the sub set of linguistic terms which appears in summary $Z_f$.

$$D_Q(Z_f) = \sum_{A_k \in Q} \left| L_{A_k} \cap I_{Z_f} \right| \qquad (8)$$

**Query substitution**

Generally, our method produces several alternative answers, more or less close to the original. These queries can be ordered according to the distance to the original one.

To evaluate our query repair method, we will use the food consumption table presented in table 7, since the Employee database does not contain null queries.

Let consider a relation $R = (Id, Age, Candy, Dairy\text{-}product, Lipid)$ from an FDB Food consumption table.

*Table 7. Food consumption table*

| Id | Age | Foods | | |
|---|---|---|---|---|
| | | *Candy* | *Dairy product* | *Lipid* |
| T1 | Young | Great | Average | Great |
| T2 | Child | Great | Excessive | Average |
| T3 | Adult | Low | Average | Great |
| T4 | Old | Low | Excessive | Low |
| T5 | Young | Great | Low | Great |
| T6 | Adult | Low | Great | Great |
| T7 | Old | Low | Great | Low |

| T8  | Child | Excessive | Excessive | Average |
|-----|-------|-----------|-----------|---------|
| T9  | Adult | Low       | Average   | Great   |
| T10 | Child | Excessive | Excessive | Great   |

The summary concept hierarchy presented in table 8 (APPENDIX 1) is the result generated after applying our Fuzzy-SaintEtiq approach to the Food Consumption table in table 7.

Let be consider for instance query $Q_4$:

> $Q_4$: **How much is consumed of dairy product and lipid by older people who consume a large amount of candy.**
>
> **Select** 3 0.25 *Dairy-product, Lipid*
> **From** *Food-consumption*
> **Where** *Age FEQ ($Old) THOLD 0.25*
> **AND** *Candy FEQ ($Excessive) THOLD 0.25*;

This query fails $v_f(P_f(Q_2)_{Zf})$ = FALSE and the failure node is $z_{42}$. From $z_{42}$ two sons' nodes are proposed $z_{51}$ and $z_{52}$.

$z_{42} = \{\{t_5^{0,2}, t_8^{0,1}, t_{10}^{0,4}\}, \{ YA, CA, GC, EC, AL \}\}$
$z_{51} = \{\{t_5^{0,1}, t_{10}^{0,4}\}, \{ YA, CA, GL, GC, EC, AL \}\}$
$z_{52} = \{\{t_8^{0,1}, t_{10}^{0,4}\}, \{ YA, CA, GC, EC, GD, ED, AL \}\}$

Both '*Young*' and '*Child*' are the labels of attribute *Age* which appear in $z_{42}$, $z_{51}$ and $z_{52}$.

So, we notice that we can change the 'old' labels of attribute *Age* by one or both of the 'Young' and 'Child' linguistic labels.

Then from $z_{42}$ the alternative query proposed is $Q_{sub}$:

> $Q_{sub}$: **How much is consumed of dairy product and lipid by children or young people who consume a large amount of candy.**
>
> **Select** 3 0.25 *Dairy-product, Lipid*
> **From** *Food-consumption*
> **Where** *Age FEQ ($Child, $Young)*
> **AND** *Candy FEQ ($Excessive)*;

## FUTURE RESEARCH DIRECTIONS

As futures perspectives of this work, we mention to:
- Implement the extracting and querying fuzzy summaries approach.
- Extend the FSQL language to support the top-k best answers
- Identify the hierarchy generated by our Fuzzy-SaintEtiq model as an index, the querying of summary hierarchies in a DBMS is difficult but still an attractive goal. Indexes are used to accelerate operations on the data, allowing relatively fast access to the data to be processed.

## CONCLUSION

Diversification of DB applications highlighted the limitations of Relational Database Management System RDBMS particularly on the modeling plan. In fact, in the real world, we are increasingly faced with the situation where applications need to handle imprecise data and to offer a flexible querying to their users.

Several theoretical solutions have been proposed. However, the impact of this work in practice, remained negligible with the exception of a few research prototypes based on the formal model GEFRED.

In this paper, we proposed a new approach of extracting and querying fuzzy summaries. The architecture of our proposed approach is divided in two major phases: fuzzy data organization phase and querying fuzzy summaries phase.

In the fuzzy data organization phase, we proposed a new approach to extract linguistic summarization from fuzzy databases, called Fuzzy-SaintEtiq.
Fuzzy-SaintEtiq takes the FDB records and provides knowledge. The summarization act considered like a process of knowledge discovery from database.

In the querying fuzzy summaries phase, our proposed approach allows users to efficiently querying summaries and exploits the conceptual hierarchical structure, and performs a flexible matching between the summaries and the query to return the top k result.

From an algorithmic point of view, the result of the comparison between the fuzzy summary and the query on the basis of linguistic labels from a user-defined vocabulary determines whether the fuzzy summary is a result but also whether a part of the hierarchy will be explored.

An extension to this method is also proposed. It allows answering queries that have empty result sets with semantically close data.
Our strategy is guided by the fuzzy summary hierarchy and it guarantees that results will be found for the new repaired query.

## ADDITIONAL READING

**Ines Benali-Sougui** Dr. Ines Benali Sougui (Tunisia) obtained the Engineering Master Degree in 2009 and the Ph.D. degree in electric genius in 2016, both from ENIT (Ecole Nationale d'Ingenieurs de Tunis). Her researches interest includes many aspects of fuzzy databases and flexible querying. She has authored three conference articles: A Quantitative Algorithm for Extracting Generic Basis of Fuzzy Association Rules (9th International Conference on Fuzzy Systems and Knowledge Discovery (FSKD 2012)), About Summarization in Large Fuzzy Databases (The Fifth International Conference on Advances in Databases, Knowledge, and Data Applications (DBKDA 2013)) and Flexible SQLf query based on fuzzy linguistic summaries (International Conference on Control, Engineering and Information Technology (CEIT'2013)) and two journal articles: From User Requirements to Flexible Querying of Fuzzy Summaries (International Journal of Service Science, Management, Engineering, and Technology (IJSSMET 2014)) and No-FSQL : a graph-based fuzzy NoSQL querying model (International Journal of Fuzzy System Applications (IJFSA 2016)).

**Minyar Sassi-Hidri**

## APPENDIX 1

The following abbreviations are used in the summary concept hierarchy of the Food Consumption table:
- For *Age* attribute: *CA* (Child Age), *YA* (Young Age), *AA* (Adult Age), *OA* (Old Age).
- For *Candy* attribute: *LC* (Low Candy), *AC* (Average Candy) and *GC* (Great Candy) and *EC* (Excessive Candy).
- For *Dairy-product* attribute: *LD* (Low Dairy-product), *AD* (Average Dairy-product), *GD* (Great Dairy-product) and *ED* (Excessive Dairy-product).
- For *Lipid* attribute: *LL* (Low Lipid), *AL* (Average Lipid), *GL* (Great Lipid) and *EL* (Excessive Lipid).

*Table 8. Fuzzy Concept summaries' generation*

| Level | Fuzzy summaries |
|---|---|
| 0 | $Z_1 = \{\{t_1^{0,7}, t_2^{0,2}, t_3^{0,4}, t_4^{0,1}, t_5^{0,1}, t_6^{0,2}, t_7^{0,2}, t_8^{0,6}, t_9^{0,3}, t_{10}^{0,4}\}, \{AL\}\}$ |
| 1 | $Z_{21} = \{\{t_1^{0,7}, t_2^{0,2}, t_5^{0,9}, t_6^{0,3}, t_8^{0,1}, t_9^{0,2}, t_{10}^{0,4}\}, \{YA, AL\}\}$<br>$Z_{22} = \{\{t_2^{0,8}, t_3^{0,6}, t_5^{0,9}, t_6^{0,8}, t_9^{0,7}, t_{10}^{0,4}\}, \{GL, AL\}\}$<br>$Z_{23} = \{\{t_2^{0,2}, t_3^{0,2}, t_4^{0,1}, t_6^{0,9}, t_7^{0,8}, t_8^{0,1}, t_9^{0,4}, t_{10}^{0,4}\}, \{GD, AL\}\}$<br>$Z_{24} = \{\{t_1^{0,2}, t_2^{0,1}, t_3^{0,2}, t_4^{0,1}, t_6^{0,3}, t_7^{0,2}, t_9^{0,4}\}, \{AC, LC\}\}$<br>$Z_{25} = \{\{t_1^{0,8}, t_3^{0,8}, t_5^{0,2}, t_6^{0,1}, t_7^{0,2}, t_9^{0,6}\}, \{AD, AL\}\}$<br>$Z_{26} = \{\{t_1^{0,3}, t_4^{0,9}, t_7^{0,8}, t_8^{0,4}\}, \{LL, AL\}\}$ |
| 2 | $Z_{31} = \{\{t_1^{0,3}, t_2^{0,2}, t_5^{0,1}, t_8^{0,1}, t_{10}^{0,4}\}, \{YA, CA, GC, AL\}\}$<br>$Z_{32} = \{\{t_2^{0,2}, t_5^{0,9}, t_6^{0,3}, t_9^{0,2}, t_{10}^{0,4}\}, \{YA, GL, AL\}\}$<br>$Z_{33} = \{\{t_2^{0,2}, t_6^{0,3}, t_8^{0,1}, t_9^{0,2}, t_{10}^{0,4}\}, \{YA, GD, AL\}\}$<br>$Z_{34} = \{\{t_1^{0,2}, t_2^{0,1}, t_6^{0,3}, t_9^{0,2}\}, \{YA, AC, AL\}\}$<br>$Z_{35} = \{\{t_2^{0,2}, t_3^{0,2}, t_6^{0,8}, t_9^{0,4}, t_{10}^{0,4}\}, \{GL, GD, AL\}\}$<br>$Z_{36} = \{\{t_2^{0,2}, t_4^{0,1}, t_8^{0,1}, t_{10}^{0,4}\}, \{ED, GD, AL\}\}$<br>$Z_{37} = \{\{t_1^{0,7}, t_5^{0,2}, t_6^{0,1}, t_9^{0,2}\}, \{YA, AD, AL\}\}$<br>$Z_{38} = \{\{t_2^{0,1}, t_3^{0,2}, t_4^{0,1}, t_6^{0,3}, t_7^{0,2}, t_9^{0,4}\}, \{GD, AC, AL\}\}$<br>$Z_{39} = \{\{t_3^{0,6}, t_5^{0,2}, t_6^{0,1}, t_9^{0,6}\}, \{GL, AD, AL\}\}$<br>$Z_{310} = \{\{t_1^{0,2}, t_3^{0,2}, t_6^{0,1}, t_7^{0,2}, t_9^{0,4}\}, \{AC, AD, AL\}\}$<br>$Z_{311} = \{\{t_1^{0,2}, t_4^{0,1}, t_7^{0,2}\}, \{AC, LL, AL\}\}$<br>$Z_{312} = \{\{t_4^{0,1}, t_7^{0,8}, t_8^{0,1}\}, \{GD, LL, AL\}\}$ |
| 3 | $Z_{41} = \{\{t_2^{0,2}, t_5^{0,1}, t_{10}^{0,4}\}, \{YA, CA, GL, GC, AL\}\}$<br>$Z_{42} = \{\{t_5^{0,2}, t_8^{0,1}, t_{10}^{0,4}\}, \{YA, CA, GC, EC, AL\}\}$<br>$Z_{43} = \{\{t_1^{0,2}, t_2^{0,1}\}, \{YA, CA, GC, AC, AL\}\}$<br>$Z_{44} = \{\{t_1^{0,3}, t_8^{0,1}\}, \{YA, CA, GC, LL, AL\}\}$<br>$Z_{45} = \{\{t_1^{0,2}, t_5^{0,1}\}, \{YA, CA, GC, AD, LD, AL\}\}$<br>$Z_{46} = \{\{t_2^{0,2}, t_8^{0,1}, t_{10}^{0,4}\}, \{YA, CA, GC, GD, ED, AL\}\}$<br>$Z_{47} = \{\{t_2^{0,2}, t_6^{0,3}, t_9^{0,2}, t_{10}^{0,4}\}, \{YA, GL, GD, AL\}\}$<br>$Z_{48} = \{\{t_5^{0,2}, t_6^{0,1}, t_9^{0,2}\}, \{YA, GL, AD, AL\}\}$<br>$Z_{49} = \{\{t_1^{0,2}, t_6^{0,1}, t_9^{0,2}\}, \{YA, AC, AD, AL\}\}$<br>$Z_{410} = \{\{t_2^{0,1}, t_3^{0,2}, t_6^{0,3}, t_9^{0,4}\}, \{GL, GD, AC, AL\}\}$<br>$Z_{411} = \{\{t_2^{0,1}, t_4^{0,1}\}, \{ED, GD, AC, AL\}\}$<br>$Z_{412} = \{\{t_3^{0,2}, t_4^{0,1}, t_6^{0,3}, t_7^{0,4}, t_9^{0,4}\}, \{GD, AC, LC, AL\}\}$<br>$Z_{413} = \{\{t_1^{0,2}, t_7^{0,2}\}, \{AC, AD, LL, AL\}\}$<br>$Z_{414} = \{\{t_4^{0,1}, t_7^{0,2}, t_8^{0,1}\}, \{ED, GD, LL, AL\}\}$ |
| 4 | $Z_{51} = \{\{t_5^{0,1}, t_{10}^{0,4}\}, \{YA, CA, GL, GC, EC, AL\}\}$<br>$Z_{52} = \{\{t_8^{0,1}, t_{10}^{0,4}\}, \{YA, CA, GC, EC, GD, ED, AL\}\}$<br>$Z_{53} = \{\{t_2^{0,2}, t_{10}^{0,4}\}, \{YA, CA, GC, GL, ED, GD, AL\}\}$<br>$Z_{54} = \{\{t_2^{0,1}, t_6^{0,3}, t_9^{0,2}\}, \{YA, GL, GD, AC, AL\}\}$<br>$Z_{55} = \{\{t_3^{0,2}, t_6^{0,1}, t_7^{0,2}, t_9^{0,4}\}, \{AD, GD, AC, LC, AL\}\}$<br>$Z_{56} = \{\{t_3^{0,1}, t_4^{0,1}, t_7^{0,4}\}, \{GD, AC, LC, OA, AL\}\}$ |
| 6 | $Z_{61} = \{\{t_3^{0,2}, t_6^{0,1}, t_9^{0,4}\}, \{AD, GD, AC, LC, GL, AL\}\}$<br>$Z_{62} = \{\{t_4^{0,1}, t_7^{0,2}\}, \{OA, LC, AC, LL, GD, AL\}\}$<br>$Z_{63} = \{\{t_3^{0,1}, t_7^{0,4}\}, \{AD, GD, AC, LC, OA, AL\}\}$ |
| 7 | $Z_{71} = \{\{t_{10}^{0,4}\}, \{YA, CA, GC, EC, GD, ED, GL, AL\}\}$<br>$Z_{72} = \{\{t_2^{0,1}\}, \{YA, CA, AC, GC, GL, GD, ED, AL\}\}$<br>$Z_{73} = \{\{t_1^{0,2}\}, \{YA, CA, GC, AC, LL, AD, LD, AL\}\}$<br>$Z_{74} = \{\{t_8^{0,1}\}, \{YA, CA, GC, EC, GD, ED, LL, AL\}\}$<br>$Z_{75} = \{\{t_5^{0,1}\}, \{CA, YA, GC, EC, AD, LD, GL, AL\}\}$<br>$Z_{76} = \{\{t_6^{0,1}, t_9^{0,2}\}, \{YA, GL, AD, AC, GD, LC, AL\}\}$<br>$Z_{77} = \{\{t_4^{0,1}\}, \{OA, LC, AC, LL, GD, ED, AL\}\}$<br>$Z_{78} = \{\{t_3^{0,1}\}, \{OA, AD, GD, AC, LC, GL, AL\}\}$<br>$Z_{79} = \{\{t_7^{0,2}\}, \{OA, LC, AC, LL, GD, AD, AL\}\}$ |

| 8 | $Z_8 = \{\{\emptyset\}, \{EL\}\}$ |